# A note on linear prediction of large chaotic systems

M.LuValle[1]

[1]OFS Laboratories

*mjl@ofsoptics.com

## Abstract

Reliable prediction of large chaotic systems in the short to middle time range is of interest in a number of fields: climate, ecology, seismology, and economics for example. In this paper, results from chaos theory and statistical theory are combined to suggest rules for building linear predictive models of chaotic systems. The rules are tested on a problem identified as hard in the climate literature, interseasonal to interannual prediction of regional seasonal precipitation. In a test of prediction, the method yields third season ahead predictions in 4 regions over 5 seasons which beat the NOAA climate prediction center's half season ahead predictions. In a test using the dimensionless climate patterns to infer parameters of the climate system, remarkable accurate estimates of rise in average global surface air temperature are produced.

**(150 character summary)**A simple modification of linear methods, to help in modeling large chaotic systems is described and applied to a "hard" prediction problem from the climate literature.

In the past decade, the combined effects of flood, and drought has resulted in the loss of thousands of lives and billions of dollars. Prediction of precipitation seasons ahead could significantly reduce these losses through providing time for preparation. However, the evolution of climate is thought to be chaotic (*1*), implying practical long term prediction in time is impossible. Adding to the difficulty, the climate system is non-stationary; with the energy available to move water and air as tracked by average global surface air temperature (GSAT) increasing over the last several decades(*2*). Neither purely empirical autoregression, nor global circulation models (GCM) are sufficiently accurate (*1,3,4,5*).

Here chaos theory(*6-12*) is combined with statistical notions to develop simple rules for linear prediction of large chaotic systems for short time periods. In any given system it remains to determine if the time period available for prediction is useful. Linear predictions are defined here as linear models of past behavior in the system to predict the future. Large chaotic systems are systems in which the box dimension of the attractor is quite large. The three rules can be stated simply if a little lack of precision can be tolerated. This will be made up for in the discussion following, in which derivation from both chaos theory and statistical ideas are laid out.

1. Long time periods of stationary behavior leading up to the time interval to be predicted are required for building and training the linear models
2. Rather than choosing a single best linear model from training, a large number of good, small linear models should be built.
3. For prediction, the mean of the models at each time should be compared to a model constructed based on a majority vote by the models (at each individual time period to be predicted), over a time interval immediately preceding that to be predicted. The winner should be the predictive model.

The first rule is easiest to derive: the variance of prediction of a linear model is dependent on both the number of points used to construct it, and the ratio of the range of the variables used in the training interval to the range in the prediction interval. Longer sampling improves both. Chaos theory[6] states that strange attractors are built from dynamic systems with fixed tuning parameters (like available thermal energy, surrogated in the climate system by GAST). The derivations of the ability of linear time series to provide prediction for chaotic systems (to date) require an attractor [10,11,12]. The current rise in GSAT over time implies it is necessary to find stationary distributions in the appropriate range from something other than the recent climate record. In this paper we accommodate using special runs of a global climate model [13,14], with fixed concentrations of green house gases to build an initial set of predictive models, while ground data is used to train, combine, and calibrate them.

The second rule arises for both statistical and chaos theory reasons. Embedology theory [7,8] guarantees that a delay map (multiple time series) is isomorphic to the original attractor only when the dimension of the delay map (n) is more than twice the box dimension (B) of the strange attractor with probability 1. For very large attractors this can make for a large number of coefficients, which statistical theory tells us to be hard to estimate from limited data. Typically a smaller number of coefficients makes for better prediction [15, 16] . However embedology theory goes on to say that if B<n<2B the delay map is isomorphic except for subspaces of size (2B-n)<B. Since this is of dimension smaller than B, the measure of such sets will be 0 with respect to the measure on the attractor, however in a neighborhood of one of these non isomorphic regions, one would expect predictions could go awry, in particular predictive models based on such delay maps would predict intermittently. In a practical situation one may have n<B. Although embedology does not say anything here, statistics does. As fewer coefficients are estimated, the dimension of the region over which prediction is done well shrinks, but decent prediction can still occur if one is lucky enough to capture major variables for the regression, and those fewer coefficients can be estimated better with a smaller sample size.

The third rule arises again from a combination of statistical and chaos theory reasons. If we are in a region where prediction is not intermittent for a large majority of the linear models, then the average should provide a good estimate of the statistical coefficient, however if a number of them are within a non-isomorphic region, there should be a cluster (of the remaining good models) in the vicinity of the true value. If there are a large number of models, and

    a)   they behave relatively independently with respect to the non-isomorphic regions, and

    b)   the probability of the non-isomorphic regions are small

Then there is high probability that the cluster in the region of the true value contains the majority of models. An attempt to construct a coherent theory of statistical methods for linear prediction of chaotic systems will have to take points a and b, either using them as assumptions, or better, proving them from what has already been assumed.

The fact that no GCM will perfectly model the true climate system implies that the models have to be selected, both within the GCM, and later from the collection of models generated in the GCM using ground data.

After hurricane Irene, the author was challenged by his supervisor to see if the method would have predicted the extreme precipitation New Jersey saw that summer. The first step was to develop a set of stationary attractors to use for initial models.

Figure 1 below shows GSAT traces for 6 runs of EdGCM 3.2® (*13,14*) assuming different amounts of greenhouse gases injected in 1958 and maintained at constant concentrations.   In each case, GSAT reaches roughly steady state by 2000 and although the traces show significant fluctuation, they are stochastically ordered by increasing year (concentration). The increase in GSAT since 1958 in these runs roughly approximates the rise in GSAT actually observed since 1958 (*2*).  These runs provide a good approximation to current climate conditions assuming that EdGCM 3.2 sufficiently represents climate physics. The steady state part of each of these 6 runs became an attractor estimate.

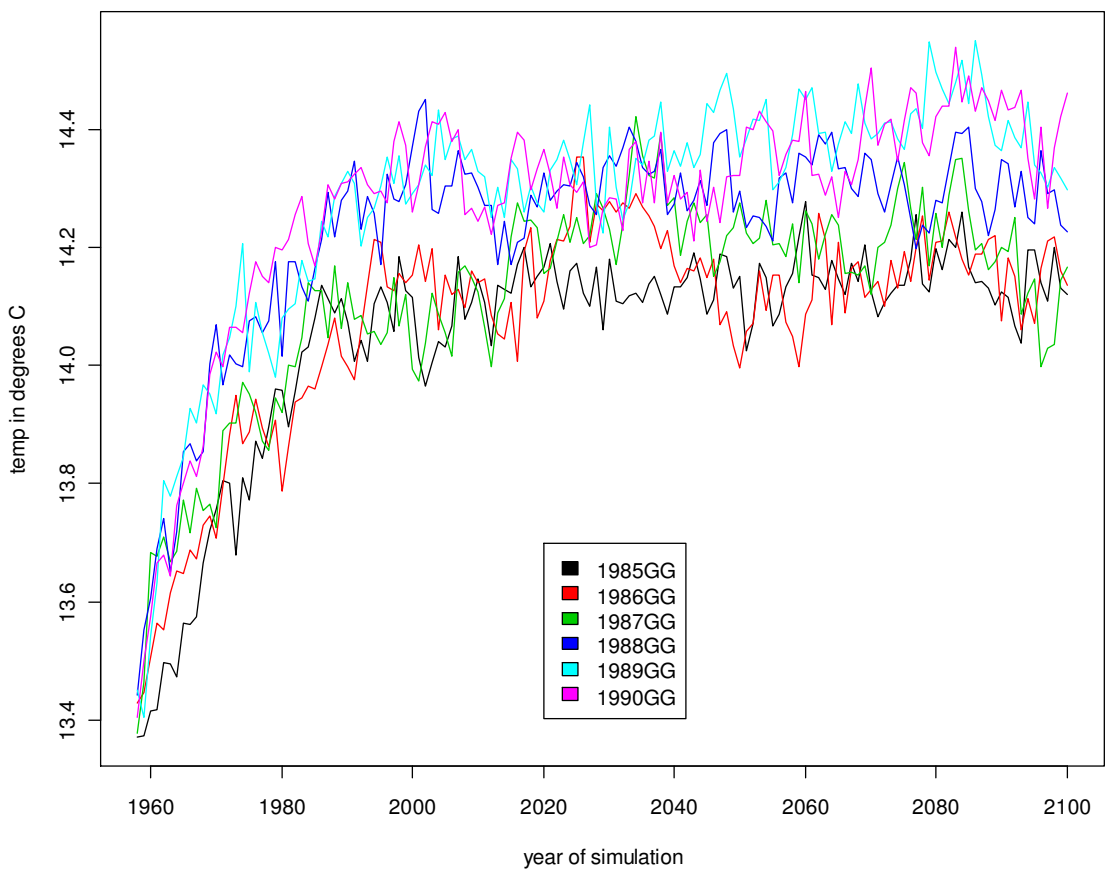

Figure 1: GSAT evolution based on EdGCM3.2® (*13,14*) simulation after injection of bolus of greenhouse gases in 1958.

Ground data was taken from individual weather stations in New Jersey and near New York City. 3rd season ahead autoregressions were developed using each of the 6 attractors ranked against 2001-2006, groups of models  were chosen by their predictions against 2007-2008 (either based on mean or clustering), and groups showing correlation better than 0.5 for 2009 and 2010 were retained (again based on mean or voting, but with possible switching).  This training data is shown as black symbols in figure 2.  The median of the set of retained predictors for each season is used to calibrate for the natural shrinkage in the predictors. The calibrated median is used for prediction. Note the precipitation in summer 2011 was strongly influence by Hurricane Irene.

The P-value for the correlation of approximately 0.894 of the 2011 prediction with the observed rainfall is 2.3e-13 using a standard student's t test for a 0 Pearsons correlation with 33 degrees of freedom. The Heidke skill (*17*), 70 in this case, is a commonly used measure of predictive skill, with 0 signifying chance prediction, 100 is perfect prediction, and -50 showing perfectly wrong prediction. The use of the standard students t test for Pearson correlation is supported by non significant results of the Box-Ljung (*18*) test for independence in each weather station precipitation series, and the fact that the predicted values use data only from before the prediction.

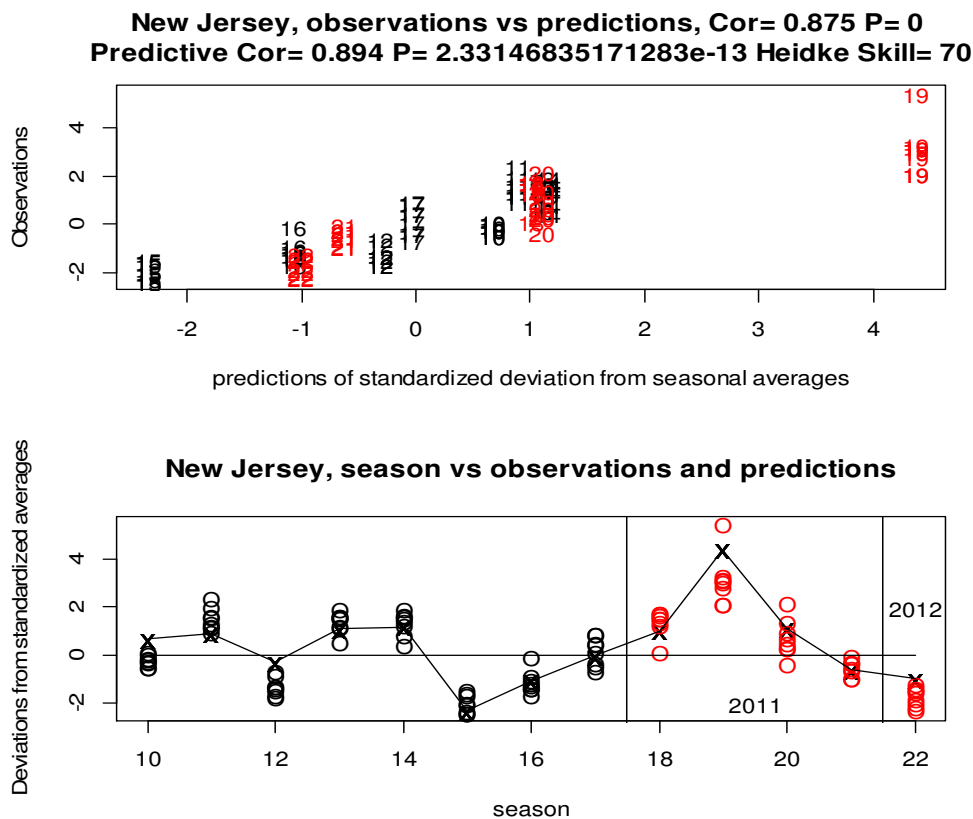

Figure 2: (2a) shows training data (black symbols) and prediction data (red symbols) of standardized variation from seasonal mean of seasonal precipitation at weather station level. The numbers in the plot show the seasons. (2b) shows a time series of this data. The line is the predictions.

In addition the method has also been tested against data in Jacksonville, FL, Atlanta GA and Southern California over the same time period with results shown in the table below.

Table 1: Correlation and Skill, Spring 2011 through spring 2012, 4 regions, 3rd season ahead prediction

| City | Predictive Pearson Correlation | P value | Degrees of freedom | Heidke Skill |
|------|-------------------------------|---------|--------------------|--------------| 
| New Jersey | 0.894 | 2.3e-13 | 33 | 70 |
| Jacksonville Fl. | 0.532 | 0.003 | 22 | 20 |
| Atlanta GA | 0.455 | 0.01 | 22 | 10 |
| Southern California | 0.757 | 6.3e-7 | 28 | 48 |

A total Heidke skill score for these regions over this time period is 56. For comparison, the 0.5 season ahead predictions for the same regions and time period by the CPC gave a Heidke skill of approximately 12.

A longer term experiment samples the behavior of this predictive method across multiple decades in Northern California. Using fit to predictions from 3 time periods, the approach was inverted to deduce global surface air temperature. In particular, the method proposed here assumes the temperature determines particular anomaly data patterns (the statistical properties of the attractor). This plot indicates that at least locally in Northern California for these three time periods (and for the training period leading up to that), the converse is true as well, providing supporting empirical evidence for the chaos driven relationships assumed here. This also demonstrates the potential of ground measured climate anomaly patterns to be used to identify better fitting attractors. Exploring other parameters of the climate model in this manner should allow direct statistical evaluation of the possible values using ground data.

In addition to the CPC comparison, The present method can also be compared to a previous survey by Lavers et al. (*4,5*)on 90 day ahead predictions averaged over 90 days for a much larger region (the continental United States). There, predictions had correlations potentially explaining between 0 and 49% of the variance in the observed regional precipitation. The method presented here shows predictions over 180 days ahead, over smaller regions with predictions explaining between 20% and 79% of the variance in the observed regional precipitation.

As a second comparison, in 2010, Westra and Sharma (*19*) estimated the upper bound of the predictability for the global climate explaining up to 38 % of the variance in precipitation, by doing a pure regression on sea surface temperatures, under a stationarity assumption, although they acknowledged the potential of local variation. They provided no approach to true interseasonal prediction.

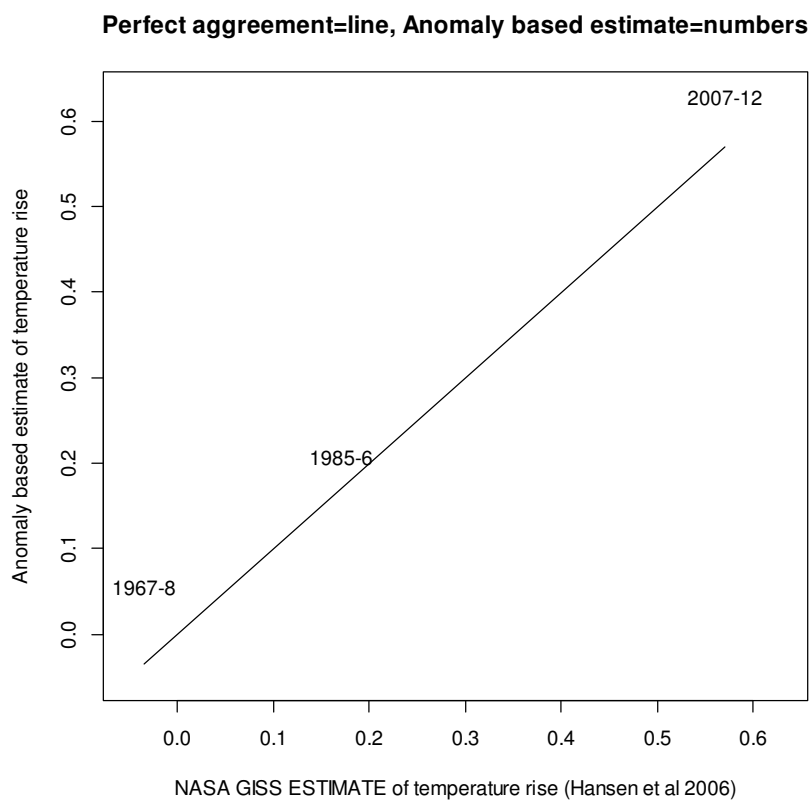

Figure 3: Rise in GSAT since 1959 (2) versus the temperature determined by the estimation procedure. The line represents perfect agreement. The year numbers plot anomaly based temperature estimates.

## Acknowledgements

I would like to thank Chris Anderson at Iowa State University for answering my early questions on climate, Dave Robbins at Rutgers for providing information on weather station data, Claudia Tebaldi at Climate Central and UBC for telling me about the fluctuation dissipation theorem, and Dave DiGiovanni for supporting my work in this area and providing significant editorial input into this paper.

Supplementary online material

Methods

A global circulation model (GCM) (EdGCM 3.2) (*12,13*) was run setting greenhouse gas concentrations to 6 different levels, and running to steady state as shown in figure 1. Then 1000 random delay maps of dimension 8 were chosen from seasonal averages of "local" precipitation and temperature, GCM approximations to the multivariate ENSO index (*20,21*), the PDO index (*22,23*), the Artic Oscillation index (*24-26*), and the North Atlantic Oscillation index (*25-27*) for 8 seasons starting at least 11 seasons prior to the season being predicted. A least squares regression was selected for each delay map using the Leaps (*28*) algorithm with the Cp (*29*) criterion for each computational cell in a neighborhood of where prediction was desired.

For each weather station these models are ordered with respect to predictive correlation against 7 years of real data. The most highly correlated X (=10%, 30% and 100%) percent of the autoregressions were extracted, and applied to predict the next two years. At this stage, both average, and average of the largest two clusters (*30*) (voting) were used for prediction. . A "key" was formed identifying the correlation, the set of cells used in the model, the lag before prediction, the percent (X) and averaging vs voting. A parametric bootstrap (*31*) helped counter the natural shrinkage caused by random predictor variables (*32*), then the individual weather stations predictions are shrunk toward the regional mean using Stein shrinkage (*15*).

The top 10 predictors (by correlation) for each attractor were tested against two further years (2009-10), predictors with correlation >0.5 were combined to predict 2011-12 using 2009-10 to calibrate.

For figure 3, attractor estimates were constructed for greenhouse gas concentrations for 17 years between 1960 and 2000. Keys were constructed for each attractor, for each time period to be predicted., False discovery rate (*33*) was applied to the keys to choose attractors to include in the temperature estimate.

An exception to this is that the analysis for Atlanta and Jacksonville does not include the NAO. An earlier draft did, showing better fit, but a computer accident resulted in some lost data, and I have not had time to recreate the full analysis. The results shown for Atlanta are from the early analysis, which I have been able to duplicate a "key" that I had. For Jacksonville I did not have a key, so was trying to guess a key to reproduce the correlation patterns in 2009, 2010. Interestingly the first two reconstructions produced enough correlation patterns satisfying the criterion, to build a prediction. Thus the results given for Jacksonville are based on two different reconstructions of the 1990 GG attractor.

To produce figure 2 and table 1, the top 10 predictors (by correlation against 2007 and 8) for each attractor were chosen. They were then tested against 2009 and 2010 together, and any predictor with correlation >0.5 was then combined using either the median (representing voting for a small number of objects) or averaging for the predictor for 2011 through spring 2012. The data were calibrated by regressing the final predictors from 2009 and 2010 against the observations.

GCM approximations to the indexes were constructing using linear combination of sea surface temperature and sea level pressure in appropriate regions for ENSO, sea surface temperature for the PDO, 1000 millibar height for AO, and 500 millibar height for the NAO (e.g. for the NAO, following L. Oman (*34*), the GCM evaluation was the difference between 2 regions both 70W to 10W, with the northern region 55 North to 70 North, the southern region was from 35 North to 45 North, this was compared to the NAO index at the NOAA website (*35*) ).

The degrees of freedom for the formal t statistic evaluation subtracts off the number of such regional seasonal means calculated. The models are rank ordered by the correlation between observations and Stein shrunken predictions when applied to real data between winter 1999 and fall 2006

Figure S.1 below shows the centers of the computational cells near the continental United States for the GCM.

Figure S.1: Centers of computational cells over the continental United States

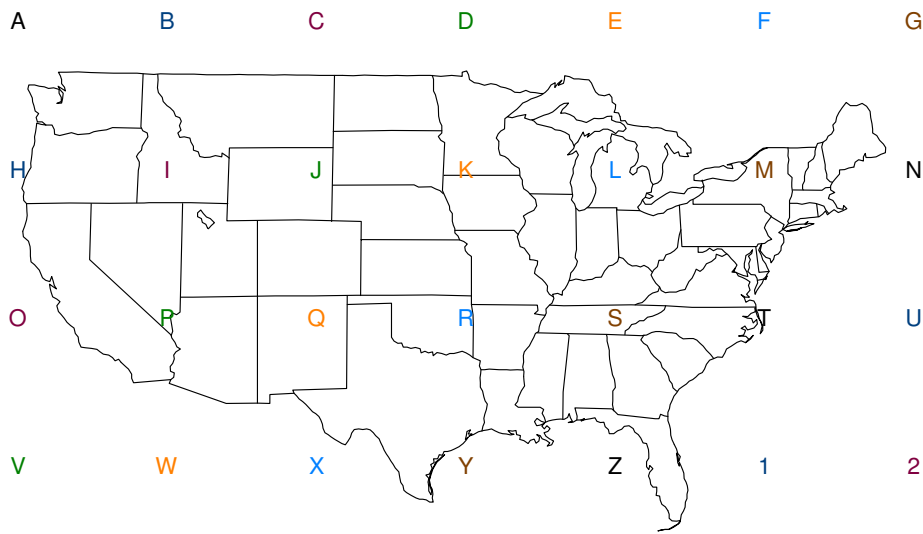

The regions used for potential of prediction in New Jersey include, M,T,1 and 2, those used in predicting Atlanta, Georgia the regions were E,L,M,R,S,T,U,Y,Z, and 1, for Jacksonville, Florida the regions were E,L,M,R,S,T,Y,Z, and 1, and for southern California the regions were H,0,P,V,W,X.

One point is worth commenting on from the numerical experiments. First there was a third method (*36*) of trying to deal with intermittent prediction examined, by directly predicting Rossby wave effects (*37*) on precipitation patterns using a time series with switching (*38*). This took significantly more effort, and our approach did not seem to provide improvement over the approaches in this paper.

Further methods used in the multi-decadal studies in Northern California

For the purpose of the multidecadal study, additional attractor estimates were constructed for greenhouse gas concentrations for 1960, 1965, 1967, 1970, 1972, 1975, 1977, 1980, 1982, 1995, and 2000. Keys were constructed for each attractor, for the time to be estimated, and years where chosen for inclusion in an average temperature estimate based on how many terms in the key would be chosen as "interesting" using a false discovery rate criterion. The plots in Figure S2 shows the False discovery rate count vs attractor year for the keys used in to construct the red numbers. The top plots show the raw counts the lower plots show the smoothed count. The chosen false discovery rate was 0.01. The temperatures were extracted from the last 100 years of each chosen run and averaged to produce the numbers plotted against the vertical axis in 3

Figure S2: FDR index index for keys for 1967-8, 1985-6, 2007-8

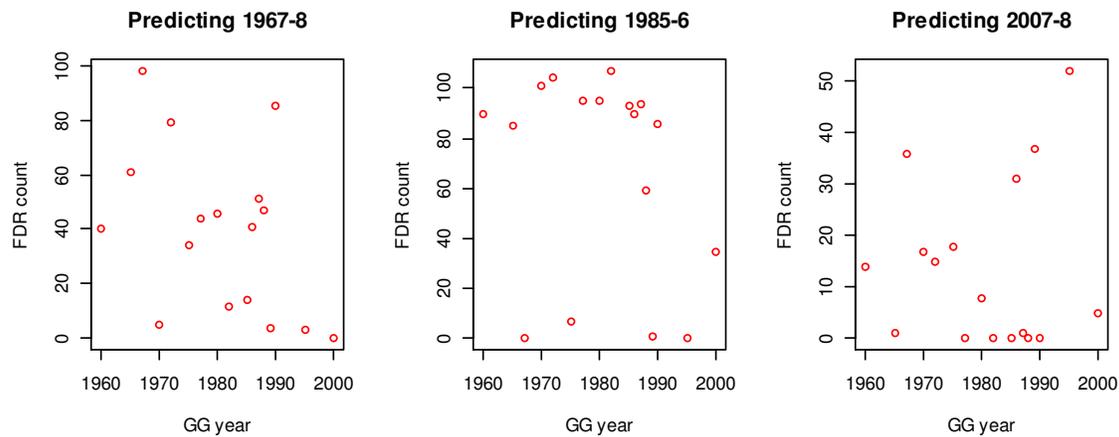

How does calculating the temperature from the anomaly time series used in building predictions support the presupposed theory used to build the precipitation predictions? The diagram below (Figure S3) describes the situation. The standard climate theory together with simple mathematics gives us all the solid black arrows, the chaos theory gives the solid blue arrows, and statistical selection gives us the red arrows. The prediction approach assumes that the GCM is close enough to the real world dynamics, that the same anomaly time series that are predictive in the GCM world are predictive in the real world. If we add the additional restrictions that the dotted arrows exist (so each mapping shown in the diagram is in fact isomorphic), then and only then will it be possible to reconstruct the global surface air temperature from the anomaly patterns. We already run into some difficulties, with the seas ice discussion above, but if we expand from temperature to include all tuning parameters. that difficulty should vanish. So the ability to deduce global temperature from anomaly patterns is consistent with and provides evidential support for the theory proposed in the body of the paper .

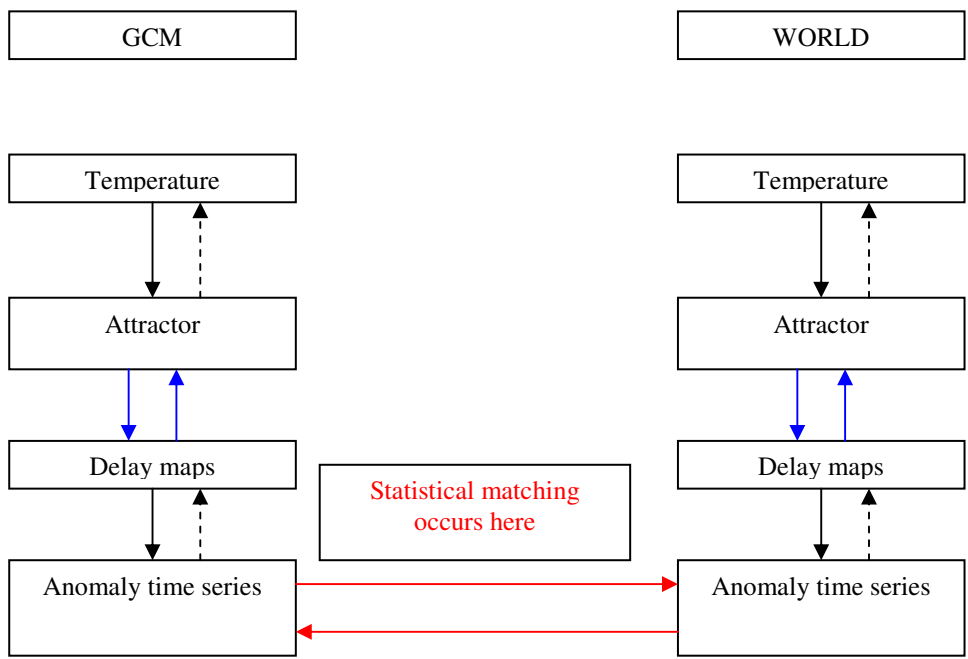

Figure S3

All predictions in for the 1967-8 and 1986-7 time frame are made between 1 year and 1.5 years ahead. The stability of the skill of prediction as lag since the initial model selection time period increased was measured using two criteria, a running Pearson correlation coefficient, and a running modified Heidke Skill score (*17*). The running statistics are calculated for the 5 weather stations over consecutive four season periods. The results are shown in figures S4 through S7 below, for two different calibration periods. Calibration here means estimating a linear regression on the original predictors constructed through the various combinations to match a prior period. The calibration periods are 8 seasons and 12 seasons. The red line in the 1990 data draws the line between statistics reflecting some data from before 1991, and those reflecting data only after winter 1990. The comparison of correlation and skill show there is still quite a bit of work to consistently do calibration well, and the comparison of the 2 calibration periods for 1990 shows how the effect of Pinatubo stands out.

Figure S4, 1970 predictions, 8 season calibration

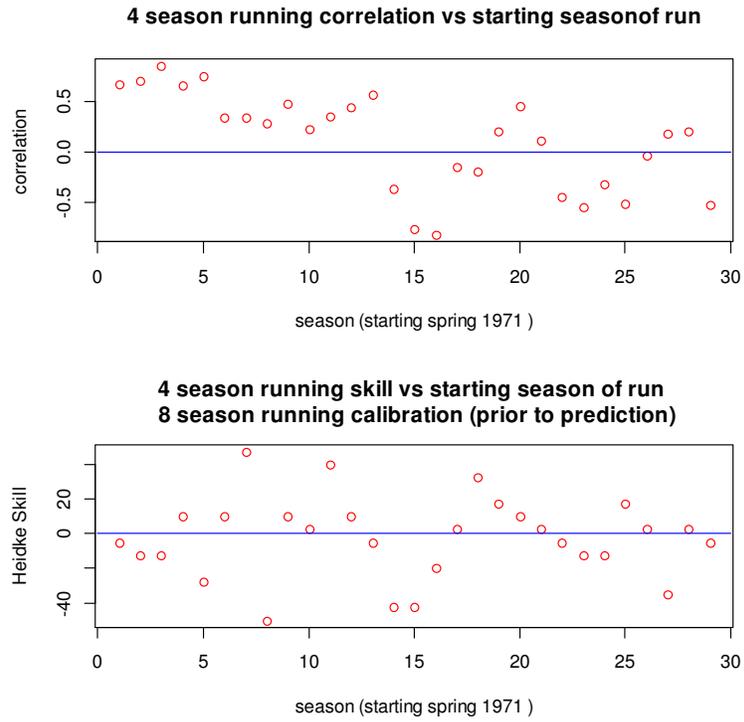

Figure S5: 1970 predictions, 12 season calibration

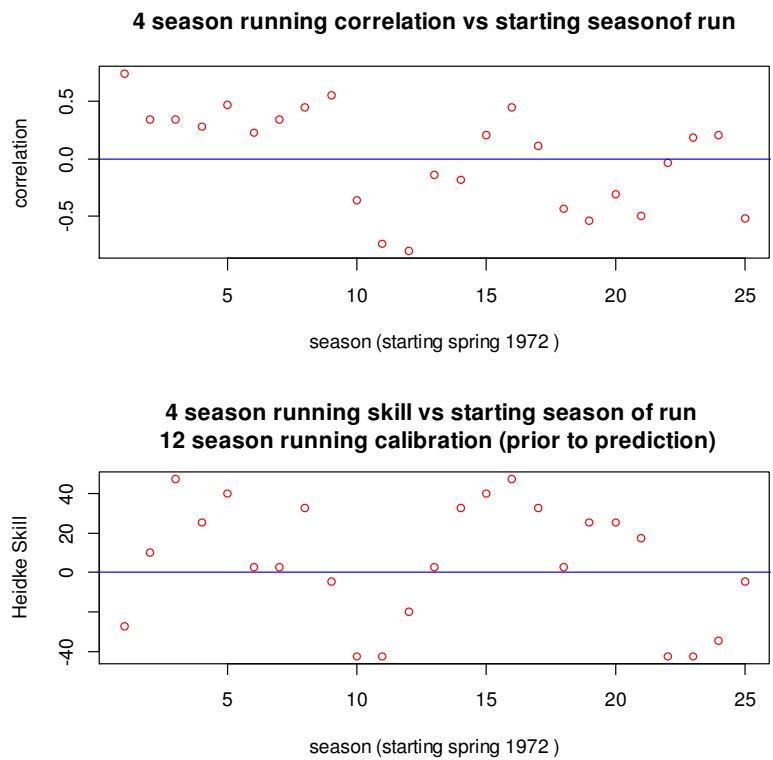

## Figure S6: 1990 period, 8 season calibration

**4 season running correlation vs starting season of run**

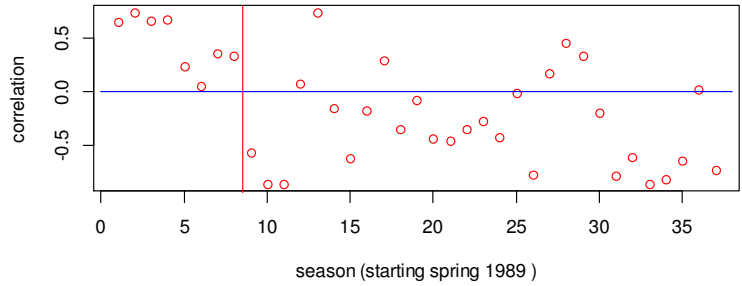

**4 season running skill vs starting season of run
8 season running calibration (prior to prediction)**

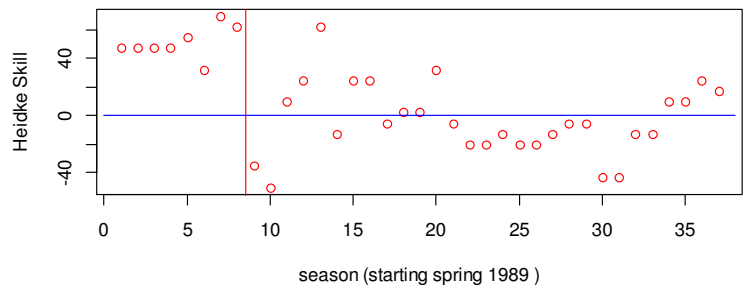

## Figure S7; 1990 12 Season calibration

**4 season running correlation vs starting season of run**

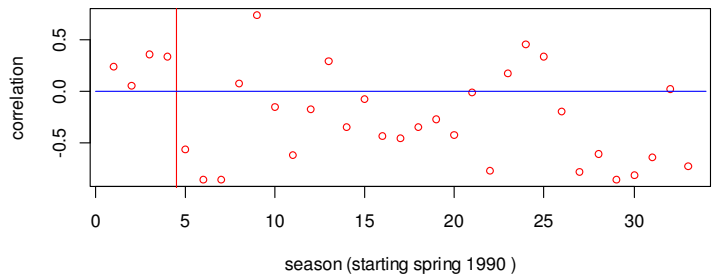

**4 season running skill vs starting season of run
12 season running calibration (prior to prediction)**

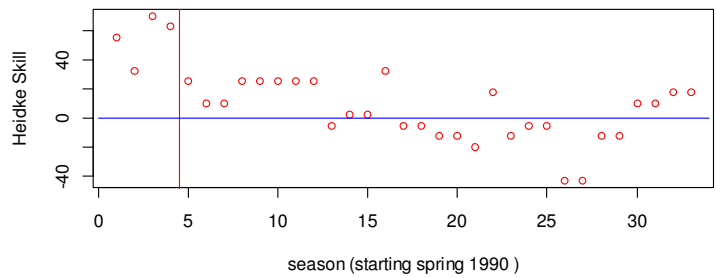

The plots show that increased calibration time increases the Heidke skill all along, but the Pearson correlation as a measure of skill drops. The later probably more accurately reflects the nonstationarity of the climate attractor because it references no real data beyond the data used to evaluate and construct the initial models. The Heidke skill includes a running calibration prior to the 4 season period being estimated.

The method used here and in the study of 4 regions differed in the following ways. First to speed up the clustering step I switched from using the kmeans algorithm to using clustering optimized for 1 dimensional splitting. Second, because of the larger number of attractors to work through, significantly less effort was spent on optimizing at the "key" formation step. Investigating the effect of the first, as well as an automated way of standardizing model selection at the "key" formation step are high priority areas for further research.

### Statistical adjustment methods

### Parametric bootstrap

When regression is performed where the predictor variables (X matrix columns) are random as well as having the dependent (Y variable) random, the coefficients are shrunken (*31*), resulting in shrunken predictions. The bootstrap (*30*) is a simple approach to estimating the sampling distribution of a statistic, even including the bias. The idea is to sample from a representative distribution in a way that represents the sampling in the original problem repeatedly. The Gaussian is a good approximation to climate variation (39), so a Gaussian model was used. However, the complicated regression model with a probabilistic combination of two time series was not attempted in this bias approximation, nor was the clustering version. Instead a simple linear regression was constructed as follows for each "weather station". The target correlation in the individual data is 0.33 corresponding to that observed in the data at that level.

1. X1=100 points were chosen independently from a standard normal distribution.
2. X2 was defined as X1+100 points chosen independently from a normal distribution with means 0 and variance 2
3. Y was defined as X1+100 points chosen independently from a normal distribution with means 0 and variance 2 (independently from those chosen for X2)
4. The regression of the first 96 Y was made on the $1^{st}$ 96 X2,
5. The last 4 Y were predicted from the last 4 X2, using the regression found on the $1^{st}$ 96.
6. The averages for the indexes 1 through 4 (pretending each index represents a season) are calculated for the full set of "weather stations" in the system.
7. This is repeated 1000s of times and the average standard deviation of the Y seasons, and predicted Y seasons were compared to estimate the expected shrinkage.

In truth, this hardly qualifies as a parametric bootstrap, as the only match attempted to the stochastic structure was in the underlying correlation at the weather station level and the number of weather stations.

### Form of the James-Stein estimate.

The estimate is made as in the original paper from the fourth Berkeley symposium[19]. Recall first the x are standardized, then when bias correction was employed, the seasonal mean is estimated by multiplying the raw data by the shrinkage bias correction from the "parametric bootstrap" calculation as described in section 3.1 giving $\hat{\mu}_{bc}$. Then the deviation $X$ of the uncorrected data from the shrinkage corrected seasonal mean is calculated and the final predicted data takes the form of:

$$\left(1 - \frac{n-2}{\|X\|^2}\right)X + \hat{\mu}_{bc} \approx \frac{\sigma_R^2}{1+\sigma_R^2}X + \hat{\mu}_{bc}$$